# Is there any measurable benefit in publishing preprints in the arXiv section Quantitative Biology?

Valeria Aman

Institute for Research Information and Quality Assurance (iFQ)
Berlin, Germany

A public preprint server such as arXiv allows authors to publish their manuscripts before submitting them to journals for peer review. It offers the chance to establish priority by making the results available upon completion. This article presents the arXiv section *Quantitative Biology* and investigates the advantages of preprint publications in terms of reception, which can be measured by means of citations. This paper focuses on the publication and citation delay, citation counts and the authors publishing their e-prints on arXiv. Moreover, the paper discusses the benefit for scientists as well as publishers. The results that are based on 12 selected journals show that submitting preprints to arXiv has become more common in the past few years, but the number of papers submitted to *Quantitative Biology* is still small and represents only a fraction of the total research output in biology. An immense advantage of arXiv is to overcome the long publication delay resulting from peer review. Although preprints are visible prior to the officially published articles, a significant citation advantage was only found for the *Journal of Theoretical Biology*.

## 1. Introduction

The arXiv is a well-established e-print server founded by Paul Ginsparg, a theoretical physicist. It was announced in 1991 by the Los Alamos National Laboratory (LANL) in New Mexico and was originally set up for the high-energy physics (HEP) community (Ginsparg, 1994). Today it is hosted by Cornell University in Ithaca, New York, and its annual budget amounts to US$750,000 (Callaway, 2013). Up to the present day it has grown outside HEP, including *Astronomy*, *Computer Science, Mathematics, Physics, Quantitative Biology, Quantitative Finance*, and *Statistics*. The number of e-prints on arXiv has grown to include 985,836 (7 Nov 2014). It receives up to 8,000 new papers each month.[1] The section *Quantitative Biology* was launched in September 2003, and is abbreviated as q-bio. Until the end of 2012 it represents 0.8% (6,456) of all preprints on arXiv.[2] *Quantitative Biology* combines techniques from biochemistry, computer science, applied mathematics and artificial intelligence.[3] In comparison to e-prints in physics, the number of biology papers is still small. Papers in the life-sciences have existed on arXiv since its launch. Nevertheless, biologists were hesitant to use arXiv, especially because they were afraid of

---

[1] arXiv (2013). arXiv submission statistics in graph form. http://arxiv.org/show_monthly_submissions
[2] arXiv submission rate statistics. http://arxiv.org/help/stats/2012_by_area/index
[3] It includes the following subject categories: Biomolecules, Genomics, Molecular Networks, Subcellular Processes, Cell Behaviour, Neurons and Cognition, Tissues and Organs, Populations and Evolution, Quantitative Methods, and Other.

getting scooped. Thus, the main users were authors from physics reporting "esoteric models and methods dabbling in biology" (Callaway, 2012). With arXiv, scientists working in one of the fields can easily disseminate their results to a wide community. Every author is welcome to upload their manuscript. There is no refereeing system but moderators who are empowered to withdraw junk papers or to move papers from one section to a more appropriate. Since its inauguration it is well-adopted by the community. Although arXiv functions without peer-review, scientist are still of the opinion that it is desirable to be finally published in a prestigious peer-reviewed journal. Therefore, a high number of papers in arXiv are simultaneously submitted for journal publication. This leads to the question, whether there is any quantifiable advantage to publish preprints on arXiv before submitting them to journals. At the same time, arXiv is also used by many authors for postprint publications, thus peer-reviewed papers that are made Open Access. The term e-print refers to both pre-and postprints and is used when no distinction is made. In general, the study investigates to what degree the arXiv is accepted among biologists as a publication platform. It zooms in on the publication delay, citation advantage, and authorship and discusses the benefits and drawbacks of preprint publications.

## 2. Data and methods

In the first step, arXiv's section *Quantitative Biology* was browsed for potential journals. For this purpose, arXiv offers the metatag "Journal-ref:", where authors can complete the information with the target publication venue of a preprint. Similarly, when authors publish their postprints on arXiv, they provide in this meta-tag the journal, where the paper was already published. Analyzing this metatag, 12 journals were chosen due to their multiple occurrences. In the next step the in-house bibliometric database with raw bibliometric information from Thomson Reuter's *Web of Science* (WoS) was applied. This database is hosted at the *Competence Centre for Bibliometrics for the German Science System*.[4] All publications of 12 journals published in the period 2003 to 2013 were aggregated. This adds up to a total of 28,201 publications. In the next step, articles from Web of Science were matched with corresponding e-prints on arXiv. The arXiv API was used to extract metadata. It is stated on the website that "the goal of the interface is to facilitate new and creative use of the vast body of material on the arXiv by providing a low barrier to entry for application developers."[5] The exact match-key was developed in the context of a master thesis and can be found on arXiv (Aman, 2013). It functions on the basis of the *PHP similar_text function* and is applied on the first author and the article title.

The main benefit of arXiv is the time advantage. Authors can upload their results as soon as they have completed their research work. In order to quantify any existing time advantage two time designations are required. Whereas arXiv provides for each e-print the date of upload, no such date exists in the database WoS. Nevertheless, our raw database with data from Thomson Reuters provides a field that indicates when an issue was loaded into the database. This data is not always precise due to delays, but overall it is a valuable approximation to the actual date of publication. The "publication delay", which rather characterizes a publication advantage, is defined as the time span between the upload of a paper on arXiv and its actual publication in a journal. In case authors upload their preprints on arXiv simultaneously to the journal submission, this time span corresponds to the length of peer-review. The duration of the peer-review-process depends heavily on the research area and is specific for a journal. Mathematical journals tend to have a longer peer-review-process than journals in other subject areas. This is due to the reason, that there is no time-pressure as in biomedicine and results once intensively proven are valid for decades.

---

[4] "Competence Centre for Bibliometrics for the German Science System" http://www.bibliometrie.info/
[5] arXiv (2012). arXiv API. http://arxiv.org/help/api/index

The publication year can easily become a fuzzy indicator for the realization of a paper. Also when it comes to citation analysis, a comparison of two publications, of which one was published in December, whereas the other was published in January in the same year, is biased towards the latter, since it had more time to accumulate citations. To have a more precise indication of the publication delay, a manual editing of the data was regarded as essential. Therefore, the date of publication was calculated on the basis of the number of volumes and issues a journal is made up of. It is sufficient to know a journal's number of volumes and issues to determine the approximate publication date. As an example, the *Journal of Theoretical Biology* publishes its issues always on the $7^{th}$ and $21^{st}$ of a month and offers very precise publication dates.

Another metric to quantify a potential benefit is the time it takes to receive the first citation. The first citation is a proof of visibility, since it guarantees that a publication was not only retrievable but also used in another paper by another author. To determine the publication date for all citing publications is an elaborate task. In this case, the loaded date in WoS was used, which is at any rate a better approximation than simply the year of publication. To overcome the bias of highly-cited reviews, the document type was restricted to articles. This is in line with the predominant document type on arXiv, which are original research articles.

## 3. Results

With the method described above, journal publications from 2003 to 2013 were matched with e-prints on arXiv. Overall, 1,034 e-prints were identified. The following table provides an overview of the number of articles found that have either a preceding preprint in arXiv or a postprint. Intended as a preprint server, the arXiv proved successful as a publication venue for peer-reviewed journal articles. More and more journals allow to place pre- and postprints on arXiv, provided that a link refers to the original journal publication.[6] Since it is of interest to reflect the broad usage of arXiv among researchers in *Quantitative Biology*, the document type was not restricted at this stage.

---

[6] Wikipedia offers a list of journals and their preprint policy:
http://en.wikipedia.org/wiki/List_of_academic_journals_by_preprint_policy

**Table 1**: Overview of biology journals, their number of papers published between 2003 and 2013, their number of e-prints matched on arXiv, and the resulting percentage.

| Rank | Journal | Nr. of papers between 2003 and 2013 | Nr. of e-prints on arXiv | % |
|---|---|---:|---:|---:|
| 1 | Journal of Theoretical Biology | 4,288 | 338 | 7.9 |
| 2 | PLoS Computational Biology | 3,056 | 183 | 6.0 |
| 3 | Physical Biology | 623 | 142 | 22.8 |
| 4 | Bulletin of Mathematical Biology | 1,083 | 101 | 9.3 |
| 5 | Journal of Mathematical Biology | 836 | 94 | 11.2 |
| 6 | Journal of Computational Biology | 1,047 | 54 | 5.2 |
| 7 | BMC Systems Biology | 1,184 | 44 | 3.7 |
| 8 | Journal of Molecular Biology | 9,379 | 34 | 0.4 |
| 9 | Journal of Evolutionary Biology | 2,388 | 16 | 0.7 |
| 10 | Journal of Biological Systems | 387 | 14 | 3.6 |
| 11 | BMC Evolutionary Biology | 2,486 | 10 | 0.4 |
| 12 | Genome Biology | 1,444 | 4 | 0.3 |
| | | **28,201** | **1,034** | **3.7** |

Source: Web of Science, arXiv and own research.

According to WoS, among these overall 1,034 papers 1,006 are genuine *Articles* (97.3%), 22 *Reviews*, 5 *Letters,* and 1 paper titled as *Editorial Material*. In the following, it will be denoted which document type is used. We can see in *Table 1* that the number of e-prints matched on arXiv differs among the journals. A journal-based investigation is meaningful on the basis of a larger dataset, so that only the first four journals from *Table 1* were chosen for more detailed analyses. Furthermore, the whole set of e-prints matched in arXiv (1,034) is compared with the set of journal publications that do not have any publication version on arXiv.

The following section presents shortly the four journals of interest. The *Journal of Theoretical Biology* (JoTB)[7] was founded in 1961 and is published by *Elsevier*.[8] It covers evolutionary biology, population genetics and immunology, with a strong focus on mathematical and computational aspects of biology. It is published biweekly on the 7$^{th}$ and 21$^{st}$ of a month. The journal's impact factor for 2013 is 2.303.[9] The journal *PLoS Computational Biology* (PCB) was only established in 2005 and is published monthly by the *Public Library of Science* (PLoS) in association with the *International Society for Computational Biology* (ISCB).[10] Although all articles are Open Access, a surprisingly high number of e-prints can be found on arXiv. The journal features articles on "living systems at all scales - from molecules and cells, to patient populations and ecosystems - through the application of computational methods".[11] Its 2013 Impact Factor is 4.829. The journal *Physical Biology* (PB) was established in 2004 and connects biology with physical sciences.[12] It is hosted by the *Institute of Physics* (IOP) where Open Access is offered at a fee of $2,700 (€1,950).[13] According to JCR the journal's Impact Factor for 2013 is 3.140. Finally, *Bulletin of Mathematical Biology* (BoMB) is an official journal of *The Society for Mathematical Biology*.[14] It is devoted to computational, mathematical, theoretical and experimental biology and addresses theorists and experimental biologists alike. Its Impact Factor for 2013 is 1.292.

---

[7] The journal's abbreviations are not official and were chosen for reasons of clarity and comprehensibility.
[8] Journal of Theoretical Biology. http://www.journals.elsevier.com/journal-of-theoretical-biology/
[9] Journal of Theoretical Biology. *2013 Journal Citation Reports*. Web of Science (Science ed.). Thomson Reuters. 2014.
[10] PLoS Computational Biology. http://journals.plos.org/ploscompbiol/
[11] PLoS Computational Biology. http://www.ploscompbiol.org/static/information

The following *Table 2* provides an overview of the publication density in order to comprehend how the date of publication was calculated. In terms of issue publication, *Physical Biology* is the journal with the least granularity. Issues are published in March, June, September and December. Thus, the publication date was set to the 15$^{th}$ of each of the enumerated months.

**Table 2**: Publication density per year for the four journals of interest.

| Journal | Nr. of volumes | Nr. of issues | Nr. of distinct dates |
|---|---|---|---|
| Journal of Theoretical Biology | *24 (since 2011)* | - | *24* |
| PLoS Computational Biology | *1* | *12* | *12* |
| Physical Biology | *1* | *4* | *4* |
| Bulletin of Mathematical Biology | *1* | *12* | *12* |

**Growth and share of papers on arXiv**

Whereas *Table 1* lists the absolute number of articles found on arXiv for the period 2003-2013, it is of interest to see developments or trends over the past years. Therefore, *Figure 1* presents on the one hand the absolute number of e-prints on arXiv per journal and on the other hand the share of e-prints in regard to a journal's total number of papers published per year. Only the four journals of interest are visualized without any restriction to the document type. The number of e-prints in *Figure 1* is presented in accordance with the year of article publication, based on the data in WoS.

*Figure 1 (a)* shows that the absolute number of preprints published on arXiv has been growing for each of the journals displayed. The steepest growth is evident for JoTB with a sudden decline in 2013. Since the number of articles published in a journal was growing over the last years, *Figure 1* (*b*) illustrates the relative share of e-prints in a journal. It is striking that PB shows a strongly fluctuating share of articles with an e-print on arXiv. The percentage varies from 15% in 2005 to 60% in 2007. This may be affected by the relative low number of articles published per year. Note that *Physical Biology* was founded in January 2004 and almost 50% of articles that constitute the first volume were published on arXiv.

---

[12] http://iopscience.iop.org/1478-3975/
[13] IOP Publishing open access policy. *http://iopscience.iop.org/info/page/openaccess*
[14] http://link.springer.com/journal/11538

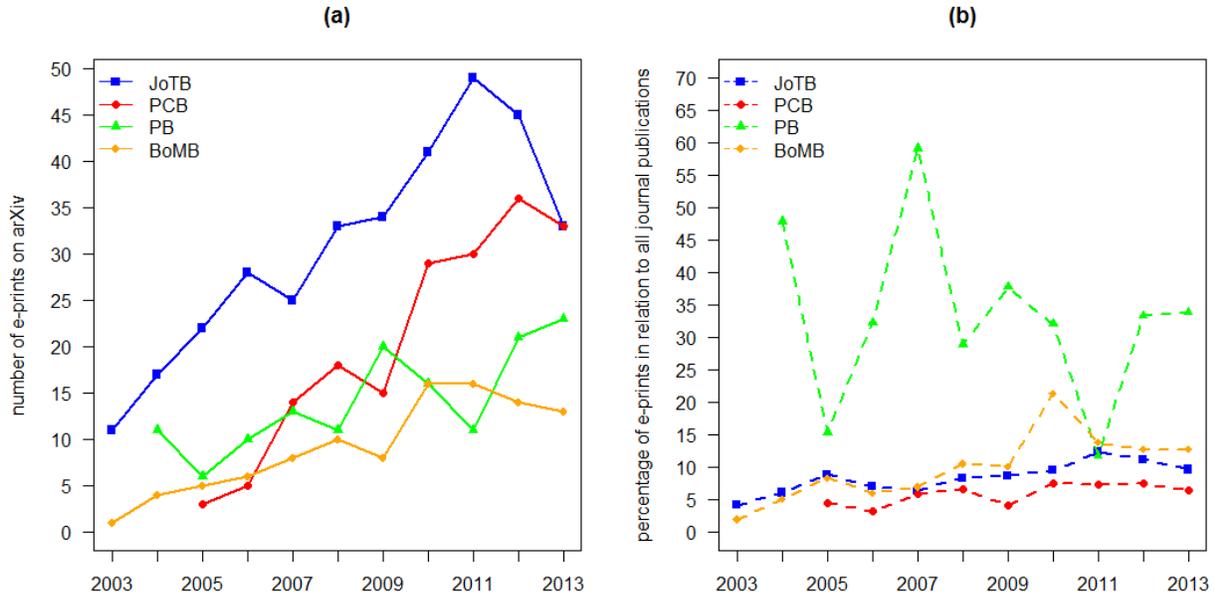

**Figure 1**: 1(a) shows the number of e-prints published on arXiv per journal. 1 (b) shows the percentage of e-prints in relation to a journal's total number of publications.

Whereas in *Figure 1* e-prints are equally displayed, the following table provides an overview of the share of genuine pre- and postprints. E-prints can be distinguished on the basis of their publication date on arXiv and the date of official journal publication. A preprint has its date of upload in arXiv prior to the date of journal publication, whereas a postprint is uploaded on arXiv after the date of journal publication. The latter was deduced from the volume and number assigned to the publication, as was explained with help of *Table 2*. Note that minimal variance has to be accepted, since the date of publication was calculated and may differ from the real date.

**Table 3**: Overview of the number and percentage of pre- and postprints on arXiv per journal.

| Journal | E-Prints | Preprints | % | Postprints | % |
|---|---|---|---|---|---|
| Journal of Theoretical Biology | 338 | 307 | 90.8 | 31 | 9.2 |
| PLoS Computational Biology | 183 | 149 | 81.4 | 34 | 18.6 |
| Physical Biology | 142 | 124 | 87.3 | 18 | 12.7 |
| Bulletin of Mathematical Biology | 101 | 93 | 92.1 | 8 | 7.9 |
| **Total** | **764** | **673** | **88.1** | **91** | **11.9** |

In *Table 3* it becomes evident that many authors publishing in *PLOS Computational Biology* place their paper on arXiv as postprints, although all peer-reviewed articles are Open Access and freely available to the community. This suggests that authors are aware of arXiv as a valuable publication repository and insist on presenting their research results there.

**Publication delay**

To determine the publication delay only genuine preprints are of interest. The type of document has not been restricted, so that articles, reviews and letters are treated equally. *Figure 2* presents the publication delay for the journals in question. To calculate the publication delay the deduced date of journal publication was considered in relation to the date of upload on arXiv. The publication delay is presented in blocks of 30 days, thus approximately a month. In order to operate with distinguishable data and to facilitate the illustration, three blocks were aggregated to a single one. In addition to the data presented in *Figure 2*, the mean and median of the publication delay in days and months is provided in *Table 4*.

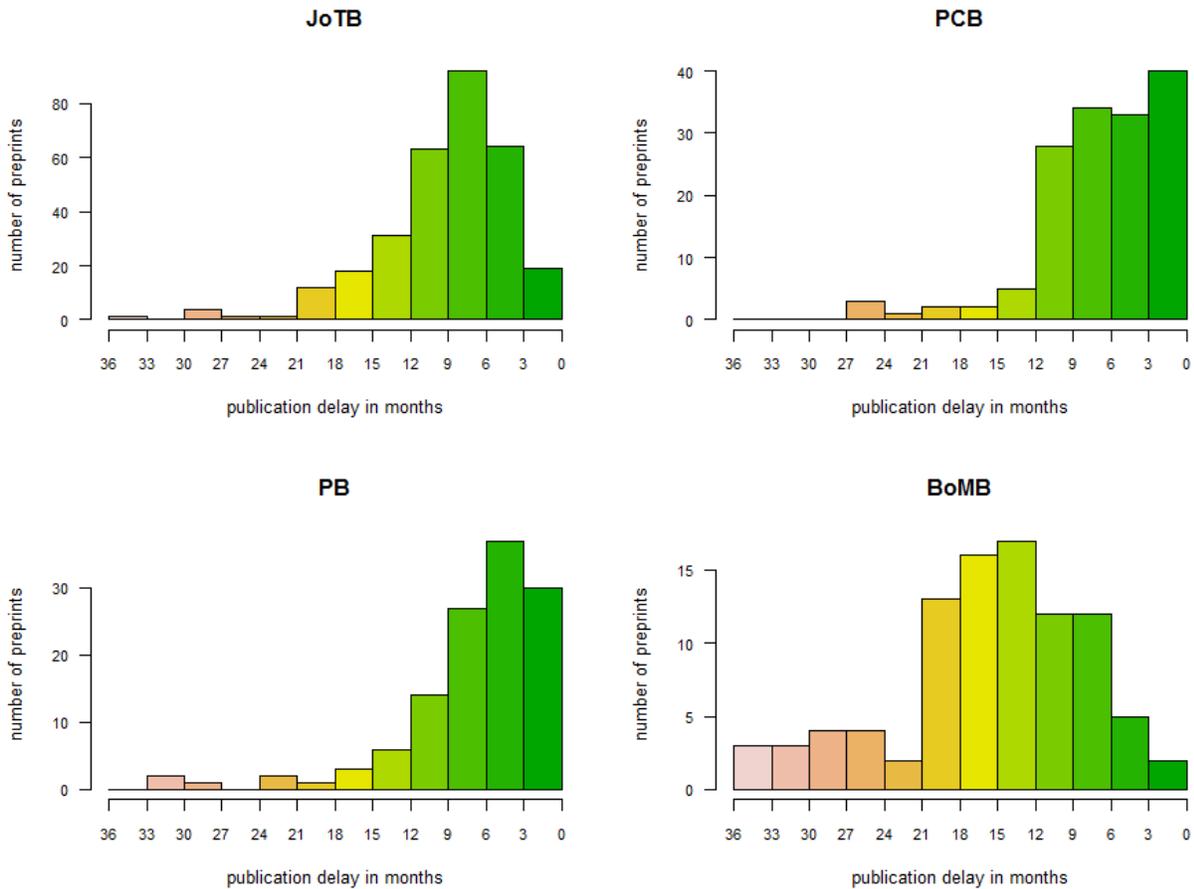

**Figure 2**: Publication delay in months aggregated to a time span of three months, thus 90 days.

For JoTB the maximum of publication delay is 6 to 9 months prior to journal publication. There is one preprint that was published prior to the publication period displayed and is thus not included in *Figure 2*. The same applies to the PB and PCB journal, each of the journals feature an outlier, thus a preprint dating back more than three years prior to the peer-reviewed publication. 82% of all preprints are placed on arXiv within one year prior to publication. The majority of preprints spend 7 to 9 months before being published.

PCB shows a less skewed distribution than JoTB and the share of preprints that are published within one year prior to publication is 92%. Only a few papers spend more than one year on arXiv before they are published in PCP. It is visible that a high share of preprints is uploaded in arXiv close to the date of publication, which may show that PCB has a relatively fast peer-review process. Note that the data presented for PCB is precise, since the publication density is high.

PB shows an extremely skewed distribution. Most of the papers are published just before the date of journal publication or may fall on the same date of publication. Due to the limited number of issues per year, this could be also an artefact of the date calculation, so that the publications in the first bloc have rather to be interpreted as postprints. This would suggest that authors publishing in this journal are more hesitant to place their papers in arXiv than authors publishing in other journals. 90% of all preprints are published within one year prior to journal publication.

Finally, for the journal BoMB the maximum of publication delay is reached at approximately 35 months, thus three years. The mode of distribution for BoMB is 16 to 18 months. 41% of all preprints are published within one year prior to journal publication. The shape of the histogram results from a long peer-review-process. 46% of all preprints are published 1 to 2 years prior to their peer-reviewed journal article. The following table sums up the results that are visible in *Figure 2*.

**Table 4**: Overview of the median and mean of publication delay in days and months.

| Journal | Mean in | | Median in | |
|---|---|---|---|---|
| | *Days* | *Months* | *Days* | *Months* |
| Bulletin of Mathematical Biology | 452 | 15 | 430 | 14 |
| Journal of Theoretical Biology | 263 | 9 | 231 | 8 |
| Physical Biology | 199 | 7 | 158 | 5 |
| PLOS Computational Biology | 190 | 6 | 168 | 6 |

The table shows that biologists follow physicists posting papers on arXiv ahead of formal publication. In regard to the length of publication delay presented above, it appears only sensible to embrace arXiv in order to publish results as early as possible.

**Citation advantage**

The question obviously arises as to whether the publication delay is of any advantage in terms of reception. The reception of a publication can be measured with means of downloads, hits, autocomplete functions of search engines, or citations in *WoS, Scopus* or *Google Scholar*. Different from Web of Science or Scopus, databases such as *Google Scholar* or *Inspire HEP* are able to track citations of preprints. Nevertheless, it can be assumed that articles with a foregoing preprint on arXiv have a citation advantage due to their open access and the longer time of visibility.

To measure the visibility, the first citation can serve as a proxy. The first citation is crucial, showing that a publication is not only retrievable but also of value for another author.

Nevertheless, reasons to cite are complex and diverse (Erikson & Erlandson, 2014). The shorter the time span between the completion of an article and its first citation, the more it is likely to be of value. In order to prove whether publications with a preprint version on arXiv are earlier cited than those without a preprint version, the two data sets have to be compared. To test for significance, Welch's t-test has been applied, which compares the means of two unpaired groups. The underlying assumption is that both datasets are samples of Gaussian distributions, but without the same standard deviation. With Welch's t-test a confidence interval for the distance of two expected values can be determined. The test presents a P-value indicating significant difference if $P < 0.05$.

To calculate the citation delay, which is defined as the time span between journal publication and the first citation received, the document type was restricted to articles. Reviews were excluded because they do not report original research results and are known for accumulating many citations, shortly after publication. To calculate the number of days to first citation, the date when an issue was loaded in WoS was used. Consequently, the time span is derived from the date of article publication and the date of the citing article, which are both based upon the date of loading. The citation window was restricted to 730 days (2 years) after the date of the published article (based upon loaded date). Only articles published between 2003 and 2011 with a fix citation window of two years were considered. Articles that were not cited during this time frame were "punished" with a length of 730 days of uncitedness. The following table lists the results for the four journals of interest. Note that postprints are in none of the sets. Only articles with a previous preprint version are compared with those articles for which neither a postprint nor a preprint exists.

**Table 5**: Overview of the time it takes to receive the first citation and the share of publications that are not cited within the first two years of publication.

| Journal | Average number of days to first citation | | Percentage of non-cited publications | |
| --- | --- | --- | --- | --- |
| | Articles without preprint | Articles with preprint | Pnc Articles without preprints | Pnc Articles with preprints |
| Journal of Theoretical Biology | 397 | 337 | 20.0 | 14.9 |
| Physical Biology | 288 | 263 | 9.1 | 8.4 |
| PLoS Computational Biology | 241 | 210 | 4.7 | 0.9 |
| Bulletin of Mathematical Biology | 411 | 427 | 26.6 | 27.4 |

From *Table 5* we can infer that the average time to receive the first citation differs among the journals listed. Whereas articles published in PCB achieve their first citation after 8 months, articles from JoTB and BoMB wait more than one year to be cited. When we have a look at the third column, we can see with exception of *Bulletin of Mathematical Biology* that articles having a previous preprint on arXiv receive their citation earlier than those without. On average, articles from PCB with a preprint on arXiv receive their first citation one month earlier and those in JoTB even two months in advance of articles without any e-print on arXiv. The indicator *Percentage of non-cited* papers (Pnc) provides the share of publications that were not cited within a fix citation window of 730 days after journal publication. Overall, the *Pnc* is compliant with the length of days until the first citation. We can see that as soon as articles have a preprint version on arXiv, they remain to a lower share uncited than articles without a preprint. Just as with the citation delay, *Bulletin of Mathematical Biology* does not show any citation advantage for articles with a preprint. A significant advantage exists nonetheless for PCB, where only a single article (0.9%) with a previous preprint remains uncited within the first 730 days after publication.

In bibliometrics, citation counts are often regarded as a proxy measure of quality in total (Seglen, 1997). Since arXiv papers can be cited, it is not uncommon to receive first citations before the official journal publication appears. Even if the results presented in *Table 5* suggest a citation advantage for articles with a previous preprint on arXiv, Welch's-t-test adds clarity. A significant citation advantage only exists for JoTB, where on average articles with a previous preprint receive their first citation 30 days ahead of articles without any e-print on arXiv (P = 0.026). This becomes also evident in the share of publications that remain uncited. Whereas every fifth paper remains uncited two years after publication, articles with a previous preprint have only a probability of 14.9% to remain uncited during the first 730 days after publication.

**Citation distribution**

The results presented so far, are based on average values and do not provide any insight into the underlying characteristics of citations. Therefore, a look on the citation distribution is necessary to specify differences between articles with a previous preprint and those without. In the following, two distinct sets are analyzed. On the one hand the articles of the 4 journals (JoTB, PB, PCB, BoMB) with a previous preprint are compared with those journal articles that have neither a preprint nor a postprint version. On the other hand, in order to give a broader view of arXiv's section *Quantitative Biology* all preprints (752) matched are compared with articles of the 12 journals from *Table 1* that have neither a pre- nor a postprint version. Different from the analysis of the citation delay, only articles published between 2003 and 2011 are presented, with a fix citation window of three years. Thus, publications published in 2011 can be cited in 2011, 2012, and 2013.

*Figure 3* shows a skew distribution of articles for the set of 4 journals and all the 12 journals. Due to the much higher number of papers, the distribution of articles without a preceding preprint is smoother than those for preprints. Whereas in the set of 4 journals a lower share of articles with a previous preprint version remains uncited, it is the opposite for the broader set of 12 journals. In the set of 4 journals, articles with a preceding preprint receive on average 6,656 citations after 3 years, whereas articles without a previous preprint reach on average a citation rate of 6,097. At the same time we can infer from *Figure 3* that articles with a previous preprint version receive on average higher citation counts within the first three years after publication than articles without a preprint. Again, the set of 12 journals does not reflect any citation advantage for articles with a previous preprint.

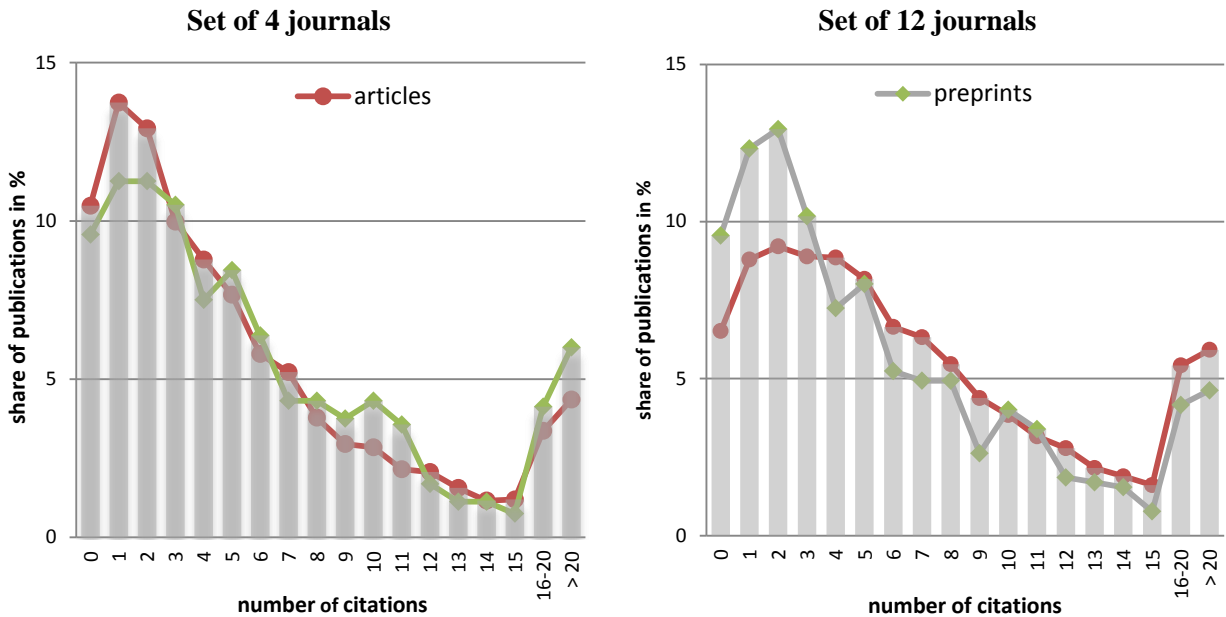

**Figure 3**: Citation distribution in percent for publications with a previous preprint on arXiv and those without. The citation window is two years and only papers published between 2003 and 2011 are displayed.

**Authors publishing in Quantitative Biology**

The bibliometric results show that there is an advantage in publishing in arXiv's section *Quantitative Biology* in order to bypass the long peer-review-process. The citation analysis shows on the other hand that articles with a previous preprint version get earlier cited and remain to a lower share uncited during the first two years after publication. This advantage is not significant, but nevertheless, many authors are aware of the benefit of arXiv and are eager to present their results online before they get published in a journal. Therefore, it is of interest to have a closer look on the authorship publishing in the section of *Quantitative Biology*.

In the first place, it is of interest to have a look on the predominant research areas. Each of the e-prints uploaded has to be assigned to subject categories within one of the seven sections on arXiv. Depending on the breadth of the paper, a multiple assignment of subject categories is possible. The following table lists the 15 most common subject categories for the set of 4 journals and the set of 12 journals. Since a multiple attribution of categories is possible and for reasons of comparability only the percentages are presented.

**Table 6**: Overview of most frequent subject categories and their share in % for the four journals of interest.

| JoTB | % | PCB | % | PB | % | BoMB | % |
|---|---|---|---|---|---|---|---|
| Populations and Evolution | 25.3 | Molecular Networks | 13.1 | Biological Physics | 19.5 | Populations and Evolution | 21.2 |
| Quantitative Methods | 9.1 | Neurons and Cognition | 12.0 | Biomolecules | 11.1 | Quantitative Methods | 11.5 |
| Biological Physics | 8.1 | Quantitative Methods | 11.4 | Subcellular Prcocesses | 11.1 | Biological Physics | 7.5 |
| Molecular Networks | 7.5 | Populations and Evolution | 11.1 | Molecular Networks | 10.5 | Molecular Networks | 6.2 |
| Statistical Mechanics | 6.9 | Biological Physics | 8.0 | Soft Condensed Matter | 9.4 | Statistical Mechanics | 4.4 |
| Cell Behavior | 6.0 | Biomolecules | 6.0 | Statistical Mechanics | 8.7 | Combinatorics | 4.4 |
| Physics and Society | 4.3 | Genomics | 4.8 | Quantitative Methods | 8.4 | Dynamical Systems | 4.4 |
| Adaptation and Self-Organizing Systems | 2.7 | Cell Behavior | 4.3 | Cell Behavior | 5.9 | Probability | 4.0 |
| Subcellular Prcocesses | 2.7 | Physics and Society | 3.4 | Populations and Evolution | 3.8 | Cell Behavior | 3.5 |
| Tissues and Organs | 2.7 | Subcellular Prcocesses | 3.1 | Genomics | 1.7 | Biomolecules | 2.7 |

The subject categories also specify the scope of the journals listed. Every fifth e-print which is published in JoTB deals with "Populations and Evolution". BoMB features the same subject category with the most often occurrence. The subject category "Biological Physics" ranks first for *Physical Biology* and reflects obviously the journal's scope. The distribution of the subject categories of e-prints that are published in *PLoS Computational Biology* display the heterogeneity of the journals' breadth of topics. The following table provides a rather holistic view of the distribution of subject categories and is based upon all 1,034 e-prints that could be matched from the 12 journals in *Table 1*.

We can infer from *Table 7* that every sixth e-print is assigned to "Populations and Evolution". Categories such as "Statistical Mechanics", "Soft Condensed Matter", or "Probability" derive from Mathematics and Physics. The table shows that the arXiv is evidently used by population biologists. As Callaway (2012) puts it, they are the ideal candidates because they are enrooted in mathematics and the open-data movement. They are aware of the fact that pre-publication brings the advantage of speed and open discussion of results. The statistical geneticist Graham Coop of the University of California said that "Biology will soon have to embrace this trend fully: the speed of discussion, comment and pre-publication review allowed is needed in biology more than most fields" (Callaway, 2012). The arXiv guarantees a wide readership and helps to establish who was first to a discovery.

**Table 7**: Overview of most frequent subject categories and their share in % for all 1,034 e-prints that could be matched. A multiple assignment of subject categories is possible.

| Rank | Subject Category | % | cum. % |
|---|---|---|---|
| 1 | Populations and Evolution | 17.3 | 17.3 |
| 2 | Quantitative Methods | 9.9 | 27.2 |
| 3 | Molecular Networks | 9.4 | 36.6 |
| 4 | Biological Physics | 8.8 | 45.4 |
| 5 | Statistical Mechanics | 5.2 | 50.5 |
| 6 | Biomolecules | 5.0 | 55.6 |
| 7 | Cell Behavior | 4.5 | 60.0 |
| 8 | Subcellular Processes | 4.0 | 64.1 |
| 9 | Neurons and Cognition | 3.1 | 67.2 |
| 10 | Genomics | 2.8 | 70.0 |
| 11 | Soft Condensed Matter | 2.7 | 72.7 |
| 12 | Physics and Society | 2.2 | 75.0 |
| 13 | Probability | 2.1 | 77.1 |
| 14 | Dynamical Systems | 1.9 | 78.9 |
| 15 | Tissues and Organs | 1.7 | 80.7 |

Another characteristic of publications is the number of participating authors. Therefore, the set of articles with an e-print version was compared with the set of articles that have neither a pre- nor a postprint on arXiv. The results are presented in *Figure 4*, split into the different sets of journals. Since it is of interest to study the willingness of authors to present their work in dependence of the number of participating authors, a distinction into pre- and postprints is insignificant.

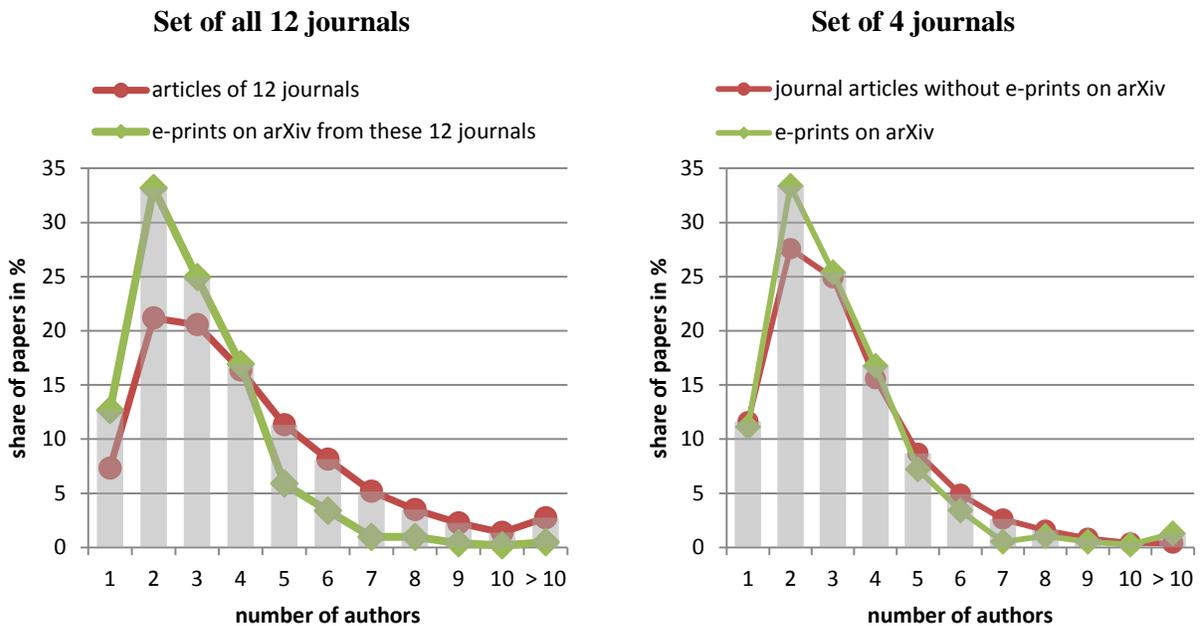

**Figure 4**: Author distribution of articles with a e-print version and those without.

From *Figure 4* we can infer that one third of publications derive from the cooperation of two authors, indifferent of the underlying journal set. In total, the distribution of authors differs remarkably in dependence of the underlying journal set. Whereas in the set of 4 journals 28% of articles without any e-print are published by two authors, the set of 12 journals shows a rate of 21%. The share of single-author-papers is similar between the two different sets of journals if an e-print exists on arXiv. A variance of share of single-author-papers is evident between the distinct journal sets, which is higher for e-prints in the set of 12 journals. It suggests that it is easier for a single author to post a preprint on arXiv and hope for useful feedback than to attain the approval of all participating authors prior to upload.

Finally, it is of interest to see the country of origin of the authors making use of arXiv's section *Quantitative Biology*. Therefore, the following table provides a ranking of countries, where most of the authors come from. Again, the whole set of 1,034 e-prints is compared with the set of e-prints matched in the four journals JoTB, PCB, PB, and BoMB.

**Table 8**: Top-15-countries for all 1,034 e-prints and those matching the 4 journals JoTB, PCB, PB and BoMB. Due to multiple participating countries, the share is indicated in %.

| Rank | Set of all e-prints matched 1,034 | | Set of e-prints from 4 journals (764) | |
|---|---|---|---|---|
| 1 | USA | 44.6 | USA | 44.2 |
| 2 | GBR | 13.9 | GBR | 15.9 |
| 3 | Germany | 11.4 | Germany | 8.4 |
| 4 | France | 8.1 | Canada | 6.4 |
| 5 | Italy | 7.2 | France | 6.2 |
| 6 | Japan | 6.5 | Japan | 5.8 |
| 7 | Spain | 4.8 | China | 5.8 |
| 8 | India | 3.7 | Spain | 3.9 |
| 9 | Canada | 3.4 | Israel | 3.7 |
| 10 | China | 3.1 | Netherlands | 3.4 |
| 11 | Israel | 3.1 | Australia | 3.4 |
| 12 | Australia | 3.0 | Italy | 3.2 |
| 13 | New Zealand | 3.0 | Switzerland | 2.8 |
| 14 | Switzerland | 2.9 | Sweden | 1.9 |
| 15 | Netherlands | 2.7 | India | 1.4 |

*Table 8* shows that USA ranks first with a lion's share of 44.6 % of all e-prints matched. It is followed by Great Britain and Germany. These three countries represent 70% of all e-prints matched. On rank 4, a different order of countries between the two sets becomes evident. Therefore, in order to see whether there are nationalities striving to publish on arXiv, the following figure displays an overlapping distribution of the most frequent countries of origin of the journals. The red curve represents the country of origin for the articles of the 12 and 4 journals respectively, whereas the green line represents the share of countries of origin for articles with an e-print version.

On the left of *Figure 5* it becomes evident that the share of US-authors using arXiv as a publication venue coincides with the authorship in the 12 journals. Authors from Germany and Great Britain are eager to post on arXiv, whereas the usage of arXiv by Canadians is below the expected value. Especially authors from Spain, Italy and India strive for publishing their papers on arXiv.

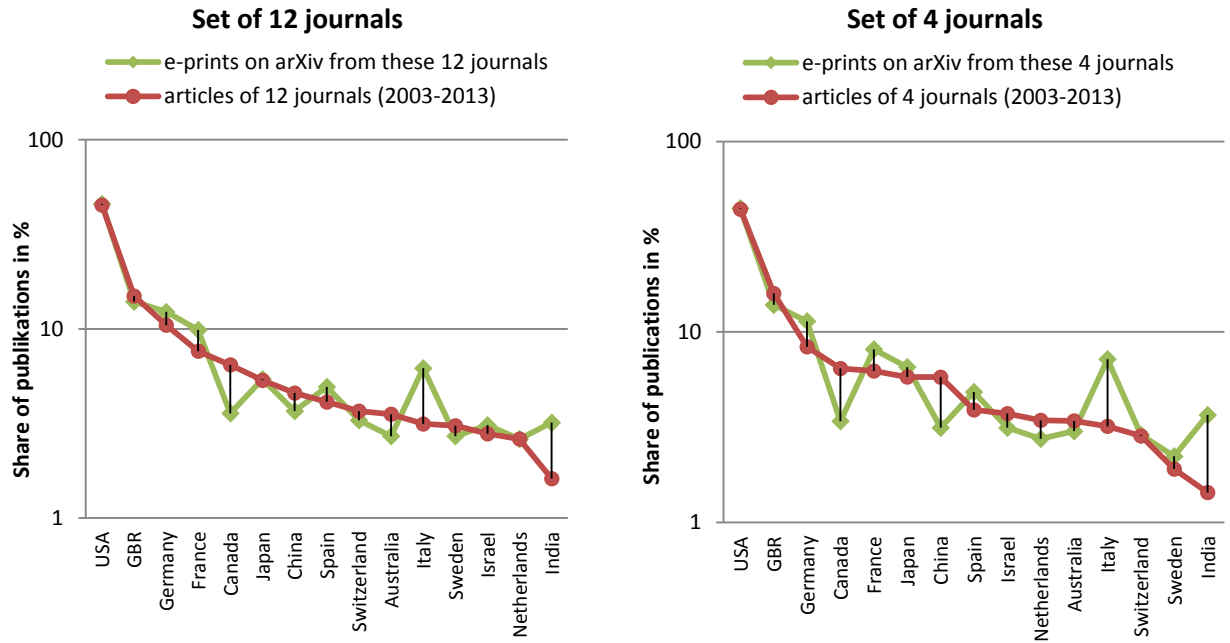

**Figure 5**: Distribution of most frequent countries using arXiv's Quantitative Biology as a publication venue. Note that the y-axis is in logarithmic scale.

The set of four journals shows almost the same distribution, but authors from France and Japan publishing in these four journals make extensively use of arXiv, whereas Chinese authors do not place their e-prints on arXiv as might be expected. Authors from Italy and India are aware of arXiv's benefit and use it more than is expected on the basis of the country of origin of the journals of interest.

## 4. Conclusion

Different from other dominating disciplines on arXiv, biology has no preprint culture, except for quantitative research such as population genetics, molecular networks or epidemiology. Desjardins-Proulx et al. (2013) write that Biologists can be prejudiced against preprint servers. They may think that a preprint server encourages stealing ideas, whereas in physics they allow the establishment of precedence and help to prevent stealing ideas. The actual idea of a preprint server is to spread new ideas. Most of the publishers are open-minded towards preprint publications, because they help to spread results without being finally formulated or omnitemporally valid. Publishers such as PLOS, Elsevier, Springer, BMC, and Nature allow for publishing pre-prints and are equally aware of the advantages as authors. The most beneficial fact is the immediacy of presenting your work to the scientific community. Authors simply have to submit the PDF. The time between submission and the final publication of a manuscript can be measured in months, as it was done in this small study. *Figure 1* showed that it can take several months before a submitted paper is published and citable. During this time, the research is only known by colleagues and reviewers. It is concealed for other researchers and cannot be discussed or followed-up by the scientific community. The benefit of a preprint server is evidently the chance to make results as early as possible available to the community to establish precedence. There are numerous additional benefits in using arXiv. The work-in-progress can be disseminated to a broad audience and arXiv helps young researchers to present their outcomes to the community without fear of invisibility. They can refer to the finalized paper, even if it was not yet officially reviewed. Every paper on arXiv receives a unique and persistent identifier, indicating the year and month of publication.

The immediacy of arXiv enables open discussion and the integration into current research. According to Callaway (2013) scientists who have studied infectious diseases used arXiv to rapidly report on ongoing outbreaks, such as the H7N9 influenza infections in China in early 2013. This applies also to the recent epidemic Ebola. The search for "Ebola"[15] leads to 13 hits, of which 8 are in the context of the 2014 Ebola epidemic. It shows that arXiv is used as a venue for up-to-date research. A benefit can be also seen on the publisher's site: Errors can be avoided before they are officially published.

Nevertheless, authors may be worried that the community does not notice their work on arXiv. This was the incentive for the creation of bioRχiv, a free online archive and distribution service for unpublished preprints in the life sciences. It was launched in November 2013 by the Cold Spring Harbor Laboratory (CSHL), a not-for-profit research and educational institution.[16] The aim is to accelerate the speed of dissemination of results and help scientists to get feedback prior to formal peer-review. Everyone is free to submit a paper, but not every paper will be published. BioRχiv offers a commenting feature, where peers can provide relevant feedback input before they are proof-read by reviewers. According to Kaiser (2013) a group of 40 "affiliate" scientists screen submissions to assure that it is real science. Paul Ginsparg himself serves on bioRχiv's advisory board.

The results of the citation analysis underpin the authors' latent worry that their work will not reach the relevant audience. It is obviously hard to see any benefit in posting papers prior to peer-review-process, if there is no guarantee for valuable comments. Due to these concerns the website *Haldane's Sieve* was created. Their aim is to "provide a simple feed of preprints in the fields of evolutionary and population genetics".[17] It features an aggregation of publications, among which those from arXiv can be found.

---

[15] arXiv. Search. http://arxiv.org/find/all/1/all:+Ebola/0/1/0/all/0/1
[16] bioRχiv. http://biorxiv.org/
[17] Haldane's Sieve. http://haldanesieve.org/about/

Nowadays, reviewers are overloaded with work, due to increasing numbers of scientists and the growing pressure to publish (Hochberg et al., 2009). Injustice exists in the different duration of a journal's peer-review process. This makes it difficult to decide who came up first with an idea. In regard to the often slow peer-review process arXiv offers an upfront way to claim priority. Desjardins-Proulx et al. (2013) state that with the peer-review process we face today, the quality of papers can be improved only at the cost of an increased load on authors as well as reviewers and a growing delay for official publication. Preprints speed up the distribution of research results and leave it to the reader's discretion whether they are groundbreaking and of value.

Some authors will remain reluctant to use preprint servers and regard them as a homogenous mass of publications, which are not filtered and for which no decision about their quality is made. But in no way do they perform worse or are low quality manuscripts. These papers are not dumped in arXiv because they could not pass the reviewer process. Thus, journals do not lose their role as validators, but an open discussion prior to official publication is valuable and should exist parallel to a handful of anonymous reviewers.